\documentclass[journal]{IEEEtran}
\usepackage{graphicx}
\usepackage{array}
\usepackage{cite}
\usepackage{amsmath,amssymb,mathrsfs}
\newtheorem{theo}{Theorem}

\begin{document}

\title{a non-iterative algorithm to estimate the modes of univariate mixtures with well-separated components}

\author{
\authorblockN { Nicolas~Paul,~\IEEEmembership{Student Member,~IEEE,}
        Luc~Fety %,~\IEEEmembership{Member,~IEEE,}
        and~Michel~Terre,~\IEEEmembership{Member,~IEEE} }
\authorblockA{
\newline
\newline
Electronic and communication department, Conservatoire National des Arts et Metiers\\
292 rue saint martin, 75003 PARIS, FRANCE\\
E-mail: nicolas.paul@cnam.fr, Tel: (33) 1 40 27 25 67, Fax: (33) 1 40 27 24 81}}% <-this % stops a space

%\markboth{submission to IEEE Signal Processing Letters SAS-STAT}{submission to IEEE Signal Processing Letters SAS-STAT}

%, %\\292 rue saint martin, 75010 Paris, FRANCE. 
%\\ E-mail: nicolas.paul@cnam.fr, Tel: (33) 1 40 27 25 67 }

%\thanks{M. Paul is with the Conservatoire National des Arts et Metiers, Laboratoire Electronique et %Communications, PARIS, FRANCE}} <= parenthèse finale de \author% pour article final

\maketitle

\begin{abstract}
This paper deals with the estimation of the modes of an univariate mixture when the number of components is known and when the component density are well separated. We propose an algorithm based on the minimization of the "kp" criterion we introduced in a previous work. In this paper we show that the global minimum of this criterion can be reached with a linear least square minimization followed by a roots finding algorithm. This is a major advantage compared to classical iterative algorithms such as K-means or EM which suffer from the potential convergence to some local extrema of the cost function they use. Our algorithm performances are finally illustrated through simulations of a five components mixture.
\end{abstract}

\begin{center} \bfseries EDICS Category: SAS-STAT \end{center}

%\pagebreak % ? (uniquement pour review ?)
%
% For peerreview papers, inserts a page break and creates the second title.
% Will be ignored for other modes.
%\IEEEpeerreviewmaketitle

%---------------------------------------------------------
% INTRODUCTION
%---------------------------------------------------------
 
\section{Introduction}

%\PARstart{I}{n} 
In this paper we focus on the estimation of the modes of an univariate mixture with a known number of components. When the mixture component belongs to a parameterized family known by the estimator (gaussian mixture case for instance), the observation estimated moments can be mapped to the mixture parameters \cite{lindsay1}. Yet, when the number of components is high, the relationships between the moments and the mixture parameters are usually too complicated to be analytically solved. Alternately, the Expectation-Maximization (EM) \cite{dempster} algorithm is the most commonly used method when the mixture densities belong to a parameterized family. It is an iterative algorithm that look for the mixture parameters that maximize the likelihood of the observations. The EM iteration consists of two steps. The Expectation step estimates the probability for each observation to come from each mixture component. During the Maximization step, these estimated probabilities are used to update the estimation of the mixture parameters. If the mixture components do not belong to any parameterized family, or if the parameterized family is not known by the estimator, the moment method and the EM algorithm do not directly apply. Yet, if the mixture components density are identical and quite separated, any clustering methods can be used to cluster the data and calculate the clusters means to reach the mixture modes. A survey of the clustering techniques can be found in \cite{berkhin}. Among them, the K-means algorithm \cite{hartigan} is one of the most popular method. It is an iterative algorithm which groups the data into K clusters in order to minimize an objective function such as the sum of point to cluster mean square Euclidean distance. K-means alternately assign each data to the closest cluster center, compute the new clusters centers and calculate the resulting cost function. A data assignment is validated only if  it decreases the overall cost function. The main drawback of K-means or EM is the potential convergence to some local extrema of the criterion they use. Some solutions consist for instance in using smart initializations (\cite{mclachlan} \cite{lindsay2} for EM, \cite{bradley} for k-means) or stochastic optimization, to become less sensitive in the initialization (\cite{celeux} \cite{pernkopf} for EM, \cite{krishna} for K-means). Another drawback of these methods is the convergence speed, which can be very slow when the number of observations is high. In this contribution, we propose a non-iterative algorithm which mainly consists in calculating the minimum of the "k-product" (kp) criterion we first introduced in \cite{norsig}. The main motivation for using such criterion is that its minimization has a global solution which can be reached by a least square optimization followed by a roots finding algorithm. The paper is organized as follows: In section 2 the observation model is presented and the criterion is defined. In section 3 the criterion global minimum is theoretically calculated. In section 4 the mode estimation algorithm is described. Section 5 presents some simulations which illustrate the algorithm performances for a 5 components mixture and conclusions are given in Section 6. 

%------------------------------------------------------------------------------------
% MODEL AND DEFINITION
%------------------------------------------------------------------------------------

\section{observation model and criterion definition}

Let $\delta$ be a discrete random variable taking its values in the set $\{ a_k \}_{k\in\{1 \cdots K\}}$ of $\mathbb{R}^K$ with probabilities $\{ \pi_k \}_{k\in\{1 \cdots K\}}$ and let $v$ be a random zero-mean variable with probability density function $g(v)$. The multimodal observation $z$ is given by: 

\begin{equation}
z=\delta+v 
\label{mixture1}
\end{equation}

\noindent We call $\textbf{a}$ the vector of the modes defined by $\textbf{a}\stackrel{\Delta}{=}(a_1,a_2 \cdots a_K)^t$. We suppose that the $a_k$ are all distincts: the probability density of $z$, $f(z)$, is then a finite mixture of $K$ identical densities  with the mixing weights $\{ \pi_k \}_{k\in\{1 \cdots K\}}$: 

\begin{equation}
f(z)=\sum_{k=1}^{K}{ \pi_k g(z-a_k) }
\label{mixture2}
\end{equation}

\noindent Let $\{ z_n \}_{n\in\{1 \cdots N\}}$ be a set of $N$ observations in $\mathbb{R}^N$. In all the following we assume that $N$ is superior to $K$ and that the number of different observations is superior to $K-1$. The kp criteria $J(\textbf{x})$ is defined by:

\begin{equation}
J: \mathbb{R}^K\rightarrow \mathbb{R}^+; \  
\textbf{x} \rightarrow \sum_{n=1}^{N}{ \prod_{k=1}^{K}{\left(z_n-x_k\right)^2} }
\label{definition_J} 
\end{equation}

\noindent This criterion has been introduced in \cite{norsig}. It is clearly positive for any vector $\textbf{x}$. The first intuitive motivation for defining this criterion is its asymptotic behavior when $v$ is null. In this case, all the observations are equal to one of the $a_k$ and therefore $J(\textbf{a})=0$. $J(\textbf{x})$ is then minimal when $\textbf{x}$ is equal to $\textbf{a}$ or any of its $K!$ permutations. The second motivation is that, in the general case, $J$ have $K!$ minima that are the $K!$ permutations of one single vector which can be reached with a linear least square solution followed by a roots finding algorithm. This is shown in section 3.

%-----------------------------------------------------------------------
% SECTION 3 KPRODUCT MINIMUM
%------------------------------------------------------------------------

\section{kp global minimum}
We first provide in section \ref{section3A} some useful definitions which are needed in section \ref{section3B} to reach the global minimum of $J$.

% QUELQUES DEFINITIONS 
% ---------------------
\subsection{Some Useful Definitions}
\label{section3A}

To any observation $z_n$ we associate the vector $\textbf{z}_n$ defined by:

\begin{equation}
\textbf{z}_n\stackrel{\Delta}{=}(z_n^{K-1}, z_n^{K-2} \cdots ,1)^t, \ \ \textbf{z}_n \in \mathbb{R}^K
\label{definition_zn}
\end{equation}

\noindent The vector $\textbf{z}$ and the Hankel matrix $\textbf{Z}$ are then respectively defined by:

\begin{equation}
\textbf{z}\stackrel{\Delta}{=}\sum_{n=1}^{N}{z_n^K \textbf{z}_n}, \ \ \textbf{z} \in \mathbb{R}^K
\label{definition_z}
\end{equation}

\begin{equation}
\textbf{Z}\stackrel{\Delta}{=}\sum_{n=1}^{N}{ \textbf{z}_n \textbf{z}_n^t }, \ \ \textbf{Z} \in \mathbb{R}^{K \times K}
\label{definition_Z}
\end{equation}

\noindent Let $\textbf{y}=(y_1,\cdots,y_K)^t$ be a vector of $\mathbb{R}^K$. We define the polynomial of order $K$ $q_\textbf{y}(\alpha)$ as:

\begin{equation}
q_{\textbf{y}}(\alpha)\stackrel{\Delta}{=} \alpha^K-\sum_{k=1}^{K}{\alpha^{K-k}y_k}
\label{definition_qy}
\end{equation}

\noindent if $\textbf{r}=(r_1,\cdots,r_K)^t$ is a vector of $\mathbb{C}^K$ containing the $K$ roots of $q_{\textbf{y}}(\alpha)$ the factorial form of $q_{\textbf{y}}(\alpha)$ is:

\begin{equation}
q_{\textbf{y}}(\alpha)=\prod_{i=1}^{K}(\alpha-r_k)
\label{qy_factorise}
\end{equation}

%\begin{multline} % multiline pour papier final (voir autre suggest ieee)
\begin{equation}
q_{\textbf{y}}(\alpha)=\alpha^K-(r_1+\cdots+r_K)\alpha^{k-1}+...\\
+(-1)^K(r_1\times r_2 \cdots \times r_K)
\label{qy_esp1}
\end{equation}
%\end{multline}
\begin{equation}
q_{\textbf{y}}(\alpha)=\alpha^K-\sum_{k=1}^{K}{\alpha^{K-k}w_k(\textbf{r})}
\label{qy_esp2}
\end{equation}

\noindent where $w_k(\textbf{r})$ is the Elementary Symmetric Polynomial (ESP) (\cite{lang}) in the variables ${r_1,\cdots,r_K}$ defined by:

\begin{equation}
w_k(\textbf{r}) \stackrel{\Delta}{=} (-1)^{k+1} \sum_{ \substack{ \{j_1,\cdots,j_k\} \in \{1\cdots K\}^k \\ j_1<\cdots<j_k \leqslant K } }{r_{j_1}.r_{j_2}\cdots.r_{j_k}}
\label{definition_wk}
\end{equation}

\noindent If we call $\textbf{w}(\textbf{r})$ the vector of ESP of $\textbf{r}$ defined by:

\begin{equation}
\textbf{w}(\textbf{r})\stackrel{\Delta}{=}(w_1(\textbf{r}),\cdots,w_K(\textbf{r}))^t
\label{definition_w}
\end{equation}

\noindent the relationship between the roots and coefficients of $q_{\textbf{y}}(\alpha)$ becomes:

\begin{equation}
\textbf{y}=\textbf{w}(\textbf{r}) \Leftrightarrow \forall k \in \{1\cdots K\} \ q_{\textbf{y}}(r_k)=0  
\label{roots_coefficients}
\end{equation}

% THEOREME ET DEMONSTRATION
%--------------------------

\subsection{The KP Minimum}
\label{section3B}

The main idea is to express $J(\textbf{x})$ as a function of $\textbf{w}(\textbf{x})$: using definitions \eqref{definition_zn} and \eqref{definition_w}, the development of each term of the sum in $J$ leads to 
$J(\textbf{x})=\sum_{n=1}^{N}{ \left({z_n^K-\textbf{z}_n^t\textbf{w}(\textbf{x})}\right)^2}$. Therefore, the minization of $J$ becomes a least square minimization in the variable $\textbf{w}(\textbf{x})$. The vector $\textbf{y}_{min}$ which minimizes $\sum_{n=1}^{N}{ \left({z_n^K-\textbf{z}_n^t\textbf{y}}\right)^2}$ can be easily obtained. Now if $\textbf{x}_{min}$ is a vector such as  $\textbf{y}_{min}=\textbf{w}(\textbf{x}_{min})$ and $\textbf{x}_{min} \in \mathbb{R}^K$, then $\textbf{x}_{min}$ is clearly a minimum of $J$. According to \eqref{roots_coefficients}, $\textbf{x}_{min}$ has to contain the $K$ roots of $q_{\textbf{y}_{min}}(\alpha)$ to have $\textbf{y}_{min}=\textbf{w}(\textbf{x}_{min})$. The difficult part is to show that these $K$ roots of $q_{\textbf{y}_{min}}(\alpha)$ are always real:

\begin{theo}
if $\textbf{y}_{min}$ is the solution of $\textbf{Z}.\textbf{y}_{min}=\textbf{z}$ (where $\textbf{Z}$ and $\textbf{z}$ have been defined in \eqref{definition_z} and \eqref{definition_Z}) and if $\textbf{x}_{min}$ is a vector containing, in any order, the $K$ roots of $q_{\textbf{y}_{min}}(\alpha)$, then $\textbf{x}_{min}$ belongs to $\mathbb{R}^K$ and $\textbf{x}_{min}$ is the global minimum of $\textbf{J}$.
\end{theo}

%\noindent \textbf{Proof}:
\begin{proof}
\noindent Let \textbf{$F$} be the function defined by:
\begin{equation}
F: \mathbb{C}^K\rightarrow \mathbb{R}^+: 
\textbf{x} \rightarrow \sum_{n=1}^{N}{ \prod_{k=1}^{K}{||z_n-x_k||_{\mathbb{C}}^2} }
\label{definition_F} 
\end{equation}

\noindent The restriction of $F$ to $\mathbb{R}^K$ is the function $J$ since the observations $z_n$ are real:
\begin{equation}
\forall \textbf{x} \in \mathbb{R}^K: \ \ \ F(\textbf{x})=J(\textbf{x})
\label{restrictionRK}
\end{equation}

\noindent Now let $H$ be the function defined by:

\begin{equation}
H: \mathbb{C}^K\rightarrow \mathbb{R}^+: 
\textbf{y} \rightarrow \sum_{n=1}^{N}{\left\|z_n^K-\textbf{z}_n^t\textbf{y}\right\|_{\mathbb{C}}^2}
\label{definition_H} 
\end{equation}

\noindent The function $H$ applied to the ESP of a vector $\textbf{x}$ in $\mathbb{C}^k$ is equal to the function $F$ applied to $\textbf{x}$:

\begin{equation}
\forall \textbf{x} \in \mathbb{C}^K: \ \ \ 
F(\textbf{x})=\sum_{n=1}^{N}{ \left\|\prod_{k=1}^{K}{(z_n-x_k)}\right\|_{\mathbb{C}}^2 }
\label{F_SumNormProduct}
\end{equation}

\noindent developping~\eqref{F_SumNormProduct} using definition~\eqref{definition_wk} leads to:

\begin{equation}
\forall \textbf{x} \in \mathbb{C}^K: \ \ \
F(\textbf{x})=\sum_{n=1}^{N}{ \left\|{z_n^K-\sum_{k=1}^{K}{z_n^{K-k}w_k(\textbf{x})}}\right\|_{\mathbb{C}}^2 }
\end{equation}

\noindent including definitions~\eqref{definition_zn} and~\eqref{definition_w}:

\begin{equation}
\forall \textbf{x} \in \mathbb{C}^K: \ \ \
F(\textbf{x})=\sum_{n=1}^{N}{ \left\|{z_n^K-\textbf{z}_n^t\textbf{w}(\textbf{x})}\right\|_{\mathbb{C}}^2 }
\end{equation}

\begin{equation}
\forall \textbf{x} \in \mathbb{C}^K: \ \ \ F(\textbf{x})=H( \textbf{w}(\textbf{x}) )  
\label{FuIsHWu}
\end{equation}

\noindent The global minimum of $H$ is the linear least square solution $\textbf{y}_{min}$ given by:

\begin{equation}
\textbf{y}_{min}=\underset{\textbf{y} \in \mathbb{C}^K}{\text{argmin}}  \left\{\sum_{n=1}^{N}{\left\|z_n^K-\textbf{z}_n^t\textbf{y}\right\|_{\mathbb{C}}^2}\right\}
\label{argminH}
\end{equation}

\noindent developping~\eqref{argminH} using definitions~\eqref{definition_z} and~\eqref{definition_Z} and remembering that the coefficients of $\textbf{Z}$ and $\textbf{z}$ are real:

\begin{equation}
\textbf{y}_{min}=\underset{\textbf{y} \in \mathbb{C}^K}{\text{argmin}}  \left\{\textbf{y}^H\textbf{Z}\textbf{y}-2\text{Re}\{\textbf{y}^H\}\textbf{z} \right\}
\end{equation}
\begin{equation}
\textbf{Z}.\textbf{y}_{min}=\textbf{z}, \ \ \ \textbf{y}_{min} \in \mathbb{R}^K
\label{ymin_obtention}
\end{equation}

\noindent The Hankel matrix $\textbf{Z}$ is regular since the number of different observations is superior to $K-1$ \cite{shohat}. Sytem~\eqref{ymin_obtention} therefore have exactly one solution.  Since $\textbf{Z}$ belongs to $\mathbb{R}^{K \times K}$ and $\textbf{z}$ belongs to $\mathbb{R}^K$, $\textbf{y}_{min}$ belongs to $\mathbb{R}^K$. Now let $\textbf{x}_{min}$=$(x_{1,min},\cdots ,x_{K,min})^t$ be a vector containing, in any order, the $K$ (potentially complex) roots of $q_{\textbf{y}_{min}}(\alpha)$. One can show that the following holds: 

\noindent \hspace*{1cm}(i) $\textbf{x}_{min}$ is a global minimum of $F$ \\
\hspace*{1cm}(ii) $\textbf{x}_{min} \in \mathbb{R}^K$ \\
\hspace*{1cm}(iii) $\textbf{x}_{min}$ is a global minimum of $J$ 

\noindent Property (i) is a direct consequence of~\eqref{FuIsHWu}:
\begin{equation}
\forall \textbf{x} \in \mathbb{C}^K: \ \ \ F(\textbf{x})=H(\textbf{w}(\textbf{x})) 
\end{equation}
\begin{equation}
\forall \textbf{x} \in \mathbb{C}^K: \ \ \ F(\textbf{x}) \geq \text{min} \left\{H\right\}
\end{equation}
\begin{equation}
\forall \textbf{x} \in \mathbb{C}^K: \ \ \ F(\textbf{x}) \geq H(\textbf{y}_{min})
\end{equation}
\noindent According to (\ref{roots_coefficients}), $\textbf{y}_{min}=\textbf{w}(\textbf{x}_{min})$ and we have:
\begin{equation}
\forall \textbf{x} \in \mathbb{C}^K: \ \ \ F(\textbf{x}) \geq H(\textbf{w}(\textbf{x}_{min}))
\end{equation}
\begin{equation}
\forall \textbf{x} \in \mathbb{C}^K: \ \ \ F(\textbf{x}) \geq F(\textbf{x}_{min})
\end{equation}

\noindent which proves (i). Property (ii) can be shown by contradiction: if $\textbf{x}_{min}$ does not belong to $\mathbb{R}^K$, then for one of the $x_{k,min}$ we have $x_{k,min} \neq \text{Re}\{x_{k,min}\}$ and, since all the observations $z_n$ are real:
\begin{equation} 
\forall n \in \{1,\cdots,N\}: \ \ \
\left\|z_n-x_{k,min}\right\|_{\mathbb{C}} > \left\|z_n-\text{Re}\{x_{k,min}\}\right\|_{\mathbb{C}}
\end{equation}
\noindent which leads to:
\begin{equation}
F(\textbf{x}_{min})>F(\text{Re}\{\textbf{x}_{min}\}) 
\end{equation}

\noindent This is impossible since $\textbf{x}_{min}$ is a global minimum of $F$. This proves property (ii). We finally have to prove (iii): since $\textbf{x}_{min} \in \mathbb{R}^K$ we have, using~\eqref{restrictionRK}: 
\begin{equation}
F(\textbf{x}_{min})=J(\textbf{x}_{min}) 
\label{FuminIsJumin}
\end{equation}
\noindent Furthermore, according to~\eqref{restrictionRK}:
\begin{equation}
\forall \textbf{x} \in \mathbb{R}^K: \ \ \ J(\textbf{x})=F(\textbf{x})
\end{equation}
\begin{equation} 
\forall \textbf{x} \in \mathbb{R}^K: \ \ \ J(\textbf{x}) \geq \text{min}\{F\}
\end{equation}
\noindent then, according to property (i):
\begin{equation}
\forall \textbf{x} \in \mathbb{R}^K: \ \ \ J(\textbf{x}) \geq F(\textbf{x}_{min}) 
\end{equation}
\noindent using~\eqref{FuminIsJumin}:
\begin{equation}
\forall \textbf{x} \in \mathbb{R}^K: \ \ \ J(\textbf{x}) \geq J(\textbf{x}_{min})
\end{equation}
\noindent which proves (iii). Properties (ii) and (iii) directly lead to theorem 1.
\end{proof}

%
% MODE ESTIMATION ALGORITHM
%
\section{modes estimation algorithm}

The mode estimation algorithm consists of two steps. In the first step, the minimum of $J$, $\{x_{1,min},...,x_{K,min}\}$, is calculated, giving a first raw estimation of the set of modes. In the second step, each observation $z_n$ is assigned to the nearest estimated mode, $K$ clusters are formed, and the final set of estimated modes is given by the means of the $K$ clusters. The algorithm steps and their complexities are illustrated in table~\ref{algo}. It appears from table~\ref{algo} that the global complexity is in o$(NK+K^2)$, which is equivalent to o$(NK)$ since $N$ is superior to $K$. 

\begin{table}
\renewcommand{\arraystretch}{1.3}
\caption{kp algorithm steps and complexities}
\label{algo}
\centering
\begin{tabular} {c}
\hline
\bfseries step 1: calculate a minimum of J \\
\hline
calculate $\textbf{Z}$ and $\textbf{z}$: o$(NK)$ \\
calculate $\textbf{y}_{min}$ by solving (\ref{ymin_obtention}): o$(K^2)$ \\
calculate the roots 
$(x_{1,min}, \cdots ,x_{K,min})$ 
of $q_{\textbf{y}_{min}}(\alpha)$: o$(K^2)$ \\
\hline
\bfseries step 2: clustering and mode estimation \\
\hline
assign each $z_n$ to the closest $x_{k,min}$: o$(NK)$ \\
calculate the K means of the resulting clusters: o$(N)$ \\
\hline
\end{tabular}
\end{table}

%
% SIMULATION
%
\section{simulation}

For each simulation run, a set of $N=100$ observations is generated from the mixture described in  ~\eqref{mixture1} with $K=5$ modes, $\textbf{a}=(0,1,2,3,4)^t$ and $\pi_k=0.2$ for all $k$ in $\{1,\cdots,K\}$. The density of $v$ is a zero-mean Laplace distribution given by $g(v)=\dfrac{\lambda}{2}e^{-\lambda |v|}$ with a variance $\dfrac{2}{\lambda^2}=10^{-2}$. This leads to well separated mixture component; the observation multimodal pdf is shown in figure 1. We suppose that the form of $g(v)$ is not known by the estimator. Therefore a  moment matching method or the EM algorithms would not directly apply. The kp algorithm is then compared to the K-means algorithm. The K-means algorithm is randomly initialized and the used cost function is the point to cluster mean square Euclidean distance. K-means is stopped when the cost function no longer decreases. The number of modes is supposed to be known in each method. 10000 runs have been performed. To get rid of the permutation ambiguity, for each run $r$, the estimated mode ($\hat{\textbf{a}}_r$) accuracy is characterized by the maximal absolute distance $D_r$ between the sorted vector of mode and the sorted vector of estimated mode:
\begin{equation}
D_r\stackrel{\Delta}{=}N(\text{sort}(\textbf{a})-\text{sort}(\hat{\textbf{a}}_r))
\end{equation}

\noindent where $N(\textbf{x})\stackrel{\Delta}{=} \underset{k \in \{1\cdots K\}}{\text{max}} |x_k|$. The distribution of $D_r$ is given in figure 2. With the K-means algorithm, $D_r$ is inferior to $0.1$ for $41.3\%$ of the run and inferior to $0.2$ for $41.4\%$ of the run. Yet, for $58.6\%$ of the run, $D_r$ is superior to $0.7$, which corresponds to a poor estimation of the set of modes. In this case the K-means method has converged to a local minimum of its cost function. Typically, one estimated mode is located in the middle of two true modes (gathering two true clusters) while two other estimated modes are closed to the same true mode. In this configuration of estimated modes, $D_r$ is around $1$. On the contrary, the kp algorithm always provides an accurate set of estimated modes: $D_r$ remains inferior to $0.1$ for $98.7\%$ of the run and $D_r$ remains inferior to $0.2$ for $99.6\%$ of the run.   

\begin{figure}
\begin{center}
\includegraphics[width=1\linewidth,height=1\linewidth]{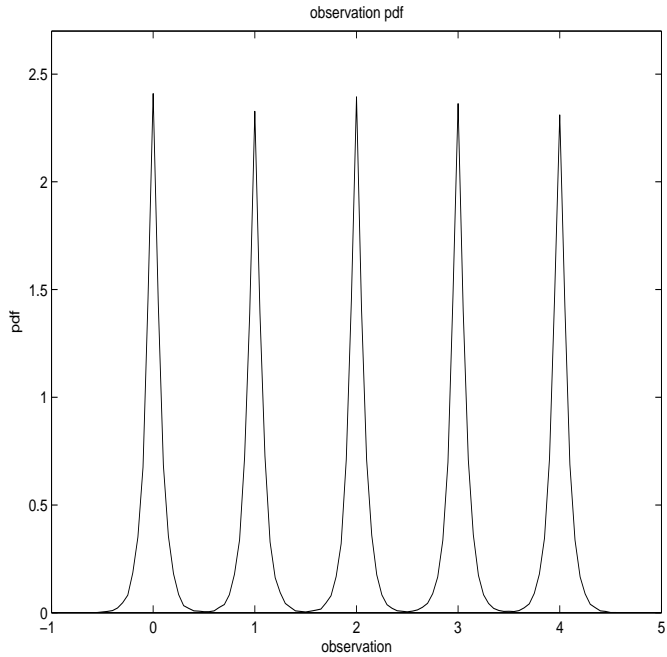}
\caption{probability density function of a five components univariate mixture with components means \{0,1,2,3,4\}, mixing weight \{0.2,0.2,0.2,0.2,0.2\} and common component density $g(v)=\dfrac{\lambda}{2} e^{-\lambda |v|}$ with variance $\dfrac{2}{\lambda^2}=10^{-2}$.}
\end{center}
\label{observationPdf}
\end{figure}

% code Danilo:
%\begin{figure}
%  \centering
%  \includegraphics{../Figures_BR/Precoder_mono.eps}
%  \caption{Estrutura}
%  \label{fig:Schema_tx_chap_mono}
%\end{figure}

\begin{figure}
\begin{center}
\includegraphics[width=1\linewidth,height=1\linewidth]{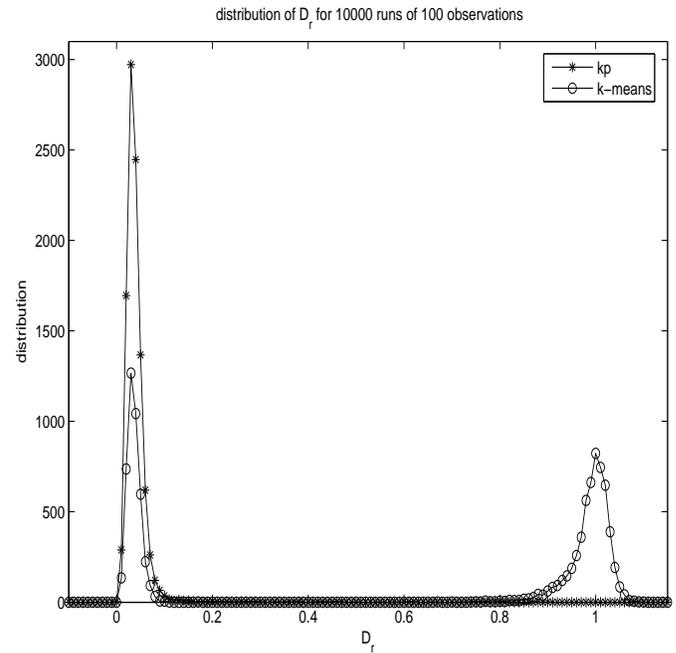}
\caption{comparison of kp (*) and k-means (o) algorithm applied to the five-component mixture described in figure 1; for each method, 10000 runs of 100 observations have been generated; for each run, the performance criteria, $D_r$, is the maximal absolute distance between the sorted vector of modes and the sorted vector of estimated modes.}
\end{center}
\label{performances}
\end{figure}

\section{conclusion}

We have provided a global minimum of the new "kp" criterion we first introduced in \cite{norsig} and used it for the estimation of the modes of univariate mixture whose component density are common and well separated. The form of the mixture densities does not have to be known by the mode estimator and does not have to belong to any particular parameterized family. Simulations have illustrated the kp algorithm good performances in the case of an univariate mixture of five Laplace distributions. In particular, the simulations have shown the superiority of our algorithm to the K-means algorithm which often converge to local minima of the used cost function. The generalization to multivariate mixture is now being studied, as well as the use of the kp criteria for estimating the number of components in a mixture.

\end{document}